\documentclass[
  journal=pasa,
  manuscript=article-type,
  year=2020,
  volume=37,
]{cup-journal}
\input{preamble}
\usepackage{lineno}
\usepackage{ulem}
\usepackage[final]{changes}
\usepackage{orcidlink}
\definechangesauthor[color=blue]{PA}
\definechangesauthor[color=purple]{DB}

\graphicspath{ {images/} }

\graphicspath{ {images/} }

\usepackage{svg}
\usepackage{float}

\title{Probing the Consistency of Cosmological Contours for Supernova Cosmology}

\author{P. Armstrong\orcidlink{https://orcid.org/0000-0003-1997-3649}}
\affiliation{Mt Stromlo Observatory, The Research School of Astronomy and Astrophysics, Australian National University, ACT 2601, Australia}
\email[P. Armstrong]{patrick.armstrong@anu.edu.au}

\author{H. Qu\orcidlink{https://orcid.org/0000-0003-1899-9791}}
\affiliation{Department of Physics and Astronomy, University of Pennsylvania, Philadelphia, PA 19104, USA}

\author{D. Brout\orcidlink{https://orcid.org/0000-0001-5201-8374}}
\affiliation{Department of Astronomy, Boston University, 725 Commonwealth Ave., Boston, MA 02215, USA}

\author{T.~M.
Davis\orcidlink{https://orcid.org/0000-0002-4213-8783}}
\affiliation{School of Mathematics and Physics, The University of Queensland, Brisbane, QLD 4072, Australia}

\author{R. Kessler\orcidlink{https://orcid.org/0000-0003-3221-0419}}
\affiliation{Kavli Institute for Cosmological Physics, University of Chicago, Chicago, IL 60637, USA}
\alsoaffiliation{Department of Astronomy and Astrophysics, University of Chicago, Chicago, IL 60637, USA}

\author{A. G. Kim\orcidlink{https://orcid.org/0000-0001-6315-8743}}
\affiliation{Physics Division, Lawrence Berkeley National Laboratory, Berkeley, CA 94720, USA}

\author{C. Lidman\orcidlink{https://orcid.org/0000-0003-1731-0497}}
\affiliation{Mt Stromlo Observatory, The Research School of Astronomy and Astrophysics, Australian National University, ACT 2601, Australia}
\alsoaffiliation{Centre for Gravitational Astrophysics, College of Science, The Australian National University, ACT 2601, Australia}

\author{M. Sako\orcidlink{https://orcid.org/0000-0003-2764-7093}}
\affiliation{Department of Physics and Astronomy, University of Pennsylvania, Philadelphia, PA 19104, USA}

\author{B. E. Tucker\orcidlink{https://orcid.org/0000-0002-4283-5159}}
\affiliation{Mt Stromlo Observatory, The Research School of Astronomy and Astrophysics, Australian National University, ACT 2601, Australia}
\alsoaffiliation{National Centre for the Public Awareness of Science, Australian National University, Canberra, ACT 2601, Australia}
\alsoaffiliation{The ARC Centre of Excellence for All-Sky Astrophysics in 3 Dimensions (ASTRO 3D)}


\begin{document}

\begin{abstract}
    As the scale of cosmological surveys increases, so does the complexity in the analyses. This complexity can often make it difficult to derive the underlying principles, necessitating statistically rigorous testing to ensure the results of an analysis are consistent and reasonable. This is particularly important in multi-probe cosmological analyses like those used in the Dark Energy Survey and the upcoming Legacy Survey of Space and Time, where accurate uncertainties are vital. In this paper, we present a statistically rigorous method to test the consistency of contours produced in these analyses, and apply this method to the Pippin cosmological pipeline used for Type Ia supernova cosmology with the Dark Energy Survey. We make use of the Neyman construction, a frequentist methodology that leverages extensive simulations to calculate confidence intervals, to perform this consistency check. A true Neyman construction is too computationally expensive for supernova cosmology, so we develop a method for approximating a Neyman construction with far fewer simulations. We find that for a simulated data-set, the 68\% contour reported by the Pippin pipeline and the 68\% confidence region produced by our approximate Neyman construction differ by less than a percent near the input cosmology, however show more significant differences \replaced{far from the input cosmology}{at the extreme extent of the 68\% contour}, with a maximal difference of 0.05 in $\Omega_{M}$, and 0.07 in $w$. This divergence is most impactful for analyses of cosmological tensions, but its impact is mitigated when combining supernovae with other cross-cutting cosmological probes, such as the Cosmic Microwave Background.%
\end{abstract}

\section{Introduction}\label{sec:introduction}
For much of the history of supernova cosmology, parameter estimation was performed via Bayesian methods which maximise the likelihood by minimising a $\chi^{2}$ function between the observed distance modulus of type Ia supernovae (SNe Ia) and the distance modulus predicted by cosmological theory.

This method of parameter estimation is shown in Equation 4 of~\cite{RiessFilippenko1998} and Equation 4 of ~\cite{PerlmutterAldering1999}, which detail the discovery of the accelerated expansion of the Universe using type Ia supernovae (SNe Ia).

Since these early efforts, parameter estimation has expanded in complexity to account for additional systematic uncertainties~\citep{ConleyGuy2011}, and to leverage large simulated data-sets to correct for contamination of core collapse supernovae~\citep{KunzHlozek2012}, and observational biases~\citep{KesslerScolnic2017}. This increased complexity is facilitated by cosmological pipelines to perform accurate parameter estimation. Due to the complex structure of modern cosmological pipelines, it is no longer possible to analytically define a likelihood function which describes the analysis being performed by these pipelines. As such, it is difficult to rigorously test the final cosmological contours produced by these cosmological pipelines, and to validate that the reported uncertainties are accurate.

There have been a number of attempts at developing alternative Bayesian frameworks which do not suffer from a non-analytic likelihood function. One such example is Approximate Bayesian Computation~\citep[ABC;][]{JenningsWolf2016, JenningsMadigan2017}, which uses realistic simulations to perform likelihood free parameter inference, at the cost of dramatically increased computation time due to the large number of simulations required. Another alternative Bayesian framework is Bayesian hierarchical models (BHM), which was implemented for supernova cosmology in~\citep[Steve;][]{HintonDavis2019},~\citep[UNITY;][]{RubinAldering2015}, and~\citep[BayeSN;][]{MandelThorp2021}. BHMs utilise multiple layers of connected parameters, allowing for a more complex analytical likelihood function to be defined.

Though these alternative frameworks have significant advantages over the $\chi^{2}$ minimisation methods, there has not been wide-spread adoption of these techniques, and many modern cosmological analyses, such as the Dark Energy Survey~\citep[DES;][]{DarkEnergySurveyCollaborationAbbott2016}, PANTHEON+~\citep{BroutScolnic2022}, and simulations of the upcoming Legacy Survey of Space and Time~\cite[LSST;][]{LSSTScienceCollaborationAbell2009,SanchezKessler2022,MitraKessler2022} still use the simpler $\chi^{2}$ method.

As modern analyses still use cosmological pipelines which rely on the $\chi^{2}$ methodology, it is important to rigorously test these cosmological pipelines. While each individual component of the SN Ia analysis pipeline is well-tested~\citep{LaskerKessler2019,KesslerNarayan2019,PopovicBrout2021,ToyWiseman2023,TaylorJones2023,KelseySullivan2023,VincenziSullivan2023}, a complete end-to-end consistency check is still necessary to understand the effects and assumptions that propagate between each individual step of the pipeline, and account for any systematic issues that may arise.

In this paper, we present a new methodology to validate the reported cosmological contours of current pipelines. For this effort, we utilize Pippin~\citep{HintonBrout2020}, which automates a number of key components of the SuperNova ANAlysis framework~\citep[SNANA;][]{KesslerBernstein2009} used in DES, LSST's Dark Energy Survey Collaboration, and PANTHEON+. Pippin and SNANA provide substantial functionality, including simulations, light-curve fitting, photometric classification training and evaluation, SNe Ia standardization and bias corrections, and cosmological fitting.

Previous efforts to construct a confidence region include those published in~\cite{BroutScolnic2019}, who used 200 simulated samples to demonstrate that the distribution of best fitting cosmologies produced by their pipeline is consistent with the average of many cosmological contours, which they took as an estimate of the confidence region (CR). This estimate of the CR, while informative, is not statistically rigorous. 

To rigorously estimate the confidence region, we make use of the Neyman construction~\citep{Neyman1937}, a Frequentist methodology that leverages simulations to produce a confidence region. The Neyman construction does not assume the CR is Gaussian or elliptical, and is robust to small sample sizes. We compare the Frequentist confidence region produced from the Neyman construction with the Bayesian contour produced by our cosmological pipeline in order to test for consistency. Producing a CR and comparing it to cosmological contours is a powerful, rigorous, and independent method of evaluating the output of a cosmological pipeline.

In Section~\ref{sec:pippin} we describe the formalism employed by the Pippin cosmological pipeline. Section~\ref{sec:methods} describes our Neyman construction methodology. Finally, the results of applying our methodology to the Pippin cosmological pipeline are presented in Section~\ref{sec:results}.%

\section{Simulation \& Analysis with Pippin}\label{sec:pippin}
Here we briefly describe the simulation of realistic data-sets of spectroscopically confirmed SN Ia, and the analysis procedure used to produce a cosmological contour. Throughout this analysis we make use of SNANA for simulation and analysis, integrated into the Pippin pipeline.

\subsection{Simulating a Supernova Data-set}
We use the SALT2~\citep{GuyAstier2007} framework within SNANA for simulating SNe Ia. \added{This framework models type Ia supernovae with five parameters: redshift, day of peak rest-frame $B$-band brightness, stretch, color, and apparent peak brightness in rest-frame $B$-band. SNANA produces simulated observed fluxes by randomly selecting these model parameters from associated probability distributions. SNANA also applies host-galaxy extinction, k-correction, and galactic extinction.}
For this analysis we use the SALT2 model produced by~\cite{TaylorLidman2021}, which was trained on a sample of 420 SNe Ia spanning a redshift range of $\sim$0.1  to $\sim$0.9 with improved zero-point calibration offsets and Milky Way extinction compared to previous SALT2 models.

\subsubsection{Bias Correction Simulations}\label{sec:biascor}
In addition to data-like simulations, we also simulate much larger data-sets to correct observational biases. As part of this analysis, we investigate the impact of the cosmology on these bias corrections. Our principal analysis uses only a single bias correction simulation with the input cosmology set to our nominal input cosmology ($\Omega_{M}=0.3$, $w=-1.0$), but we repeat our analysis using many bias correction simulations, with input cosmologies equal to the input cosmology of the data-set they are correcting, to see if this affects the Neyman construction.

\subsection{Analysis}
The supernovae in each simulated data-set are fit to determine the SALT2 parameters: amplitude ($x_{0}$), stretch ($x_{1}$), and colour ($c$). From here, the distance modulus of each SN Ia can be computed via the Tripp equation~\citep{Tripp1998}
\begin{equation}
\mu = m_{B} + \alpha{}x_{1} - \beta{}c + \mathcal{M} - \Delta\mu_{bias}
\end{equation}
Here $\alpha$ and $\beta$ are global stretch and colour nuisance parameters, $\mathcal{M}$ is a global offset, and $\Delta\mu_{bias} = \mu - \mu_{true}$ is a distance bias correction, where $\mu_{true}$ is the true distance modulus. 

Pippin makes use of the BEAMS with Bias Correction \citep[BBC;][]{KesslerScolnic2017} framework to produce a Hubble Diagram (HD) that has been corrected for both selection effects and contamination. BBC uses the detailed simulations described in Section~\ref{sec:biascor} alongside the BEAMS~\citep{KunzHlozek2012} method to correct for both distance biases, contamination, and selection effects~\citep{KesslerBrout2019}. It then uses a cosmology-independent method \citep[SALT2mu;]{MarrinerBernstein2011} to fit for global nuisance parameters and standardise the SNe Ia magnitudes.

In order to fit $\alpha$, $\beta$, and $\mathcal{M}$, BBC adopts the likelihood $\mathcal{L} = \prod_{i=1}^{N}\mathcal{L}_{i}$ where
\begin{align}
\mathcal{L}_{i} &= P_{Ia,i}D_{Ia,i} + (1 - P_{Ia,i})D_{CC,i}
\end{align}
Here $P_{Ia,i}$ is the photometric classification probability for the $i$th supernova to be an SN Ia. This is usually calculated via a photometric classifier such as SuperNNova~\citep{MollerdeBoissiere2020}, or Scone~\citep{QuSako2021}, however for our analysis we do not simulate contamination, so $P_{Ia,i}=1$. $D_{Ia,i}$ encodes the influence of SNe Ia on the likelihood, including corrections for observational biases in the data-set. Details of $D_{Ia,i}$ are presented in~\cite{KesslerScolnic2017}. The $D_{CC,i}$ component encodes the effects of contaminants; however, since our simulations are contaminant free, it is unimportant for this analysis.

The end result of the BBC framework is a redshift-binned HD. BBC can also provide an unbinned HD which~\cite{BroutHinton2021} shows can result in smaller systematic uncertainties, but is more computationally expensive. Since our analysis only includes statistical uncertainties, we gain no benefit from using an unbinned HD, therefore we only use the default, binned HD.

This binned HD is passed to a cosmological fitter to produce the final cosmological contours. In this analysis we make use of WFit, which measures a $\chi^{2}$ likelihood over a grid within the parameter space. WFit has the advantage of being much faster than other methods, although is only suitable for simple test cases such as the one used in this paper, and not necessarily suitable for final survey cosmological analysis. We allow the parameter space of $\Omega_{M}$ to vary below 0, something which is not usual for cosmological analyses, as our analysis requires WFit to explore large sections of the $\Omega_{M}$, $w$ parameter space, and we do not wish to artificially truncate the likelihood surface we produce.

\subsection{Producing an experiment data-set}\label{sec:experiment}
\begin{figure}
    \centering
    \includegraphics[width=\linewidth]{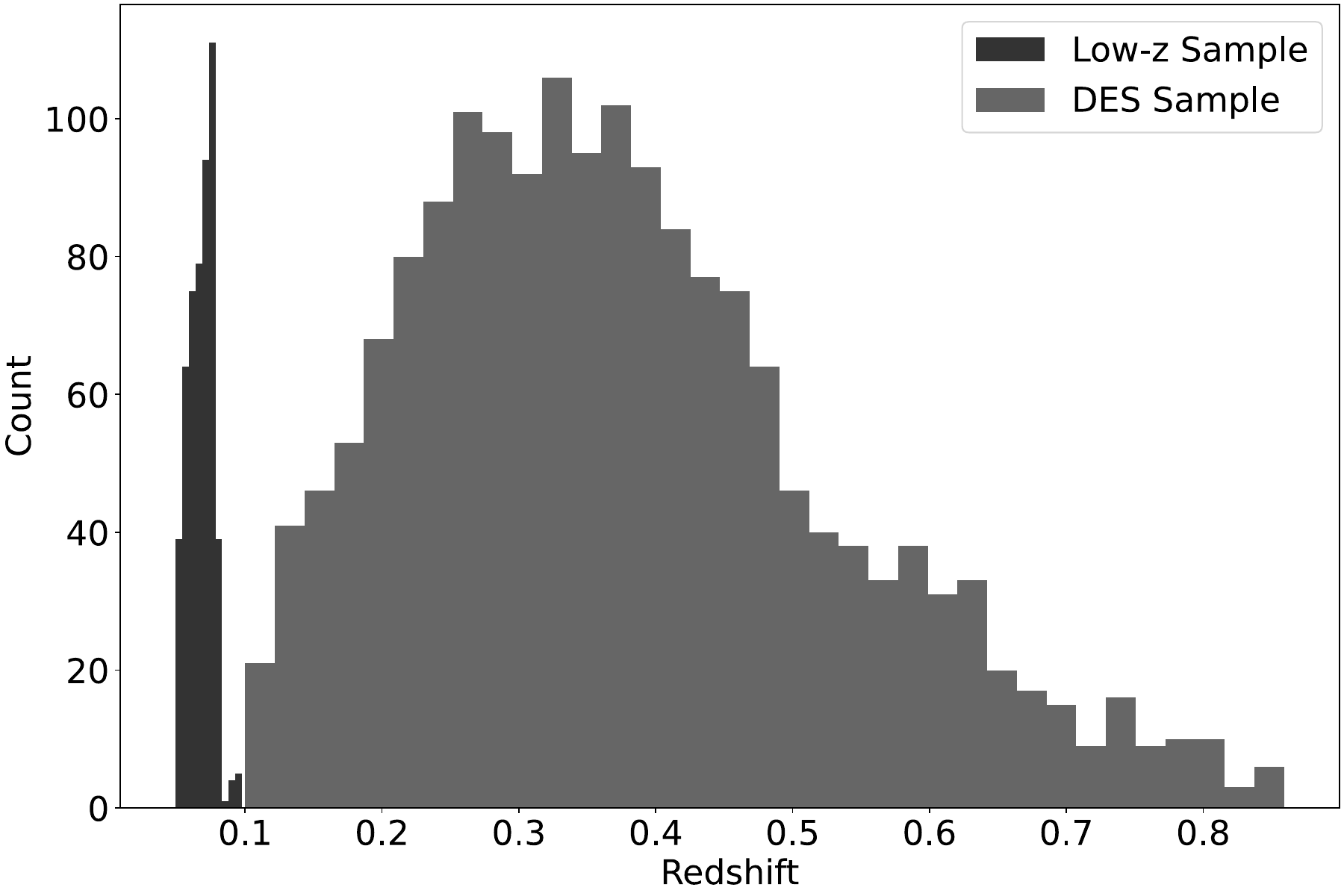}
    \caption{\added{The redshift distribution of the simulated DES sample and simulated low-$z$ sample.}}
    \label{fig:redshift}
\end{figure}

\begin{figure}
    \centering
    \includegraphics[width=\linewidth]{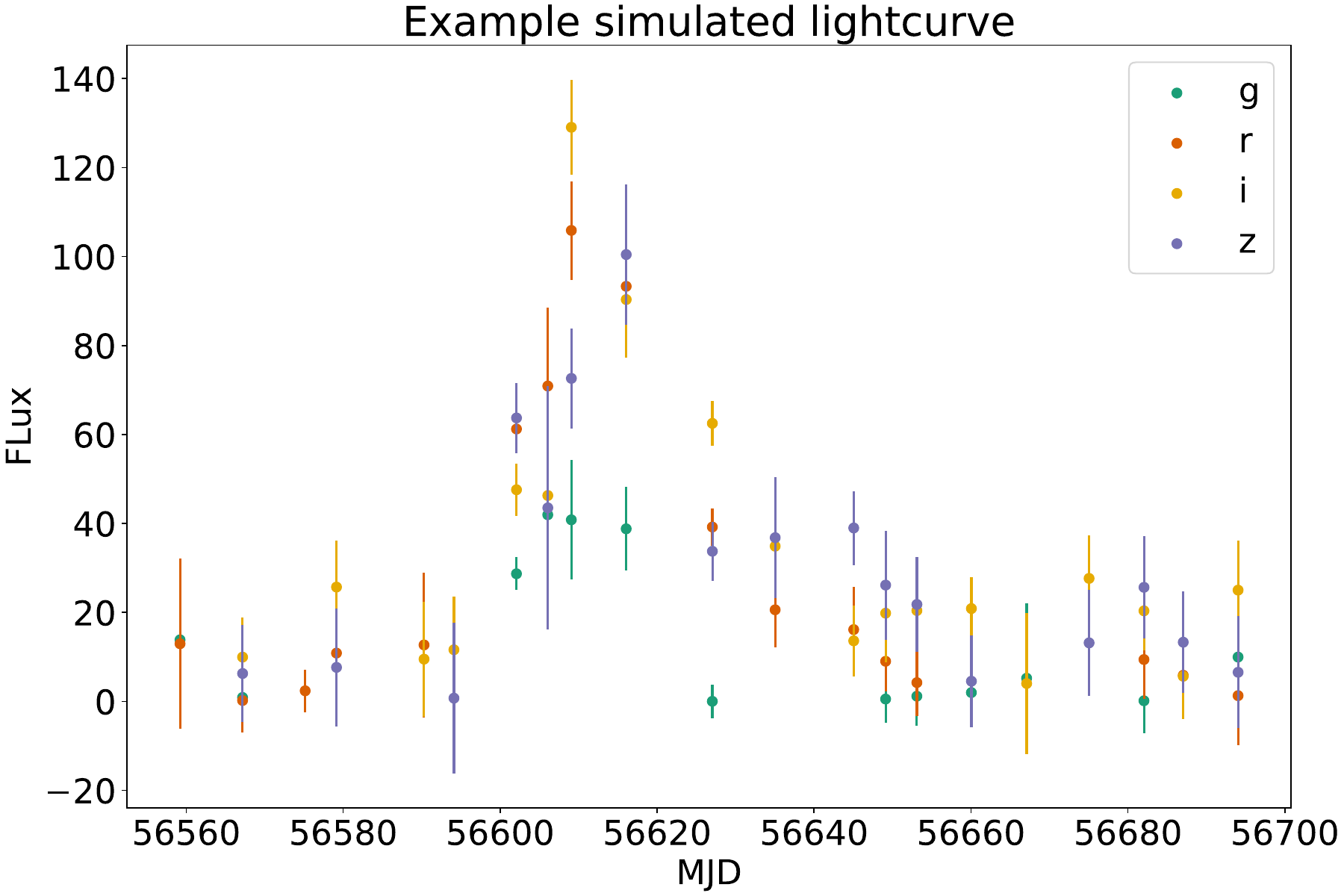}
    \caption{\added{An example of a simulated lightcurve in our simulated sample which lies at $z=0.4$.}}
    \label{fig:lightcurve}
\end{figure}

Our methodology can be used to validate the contour produced by any cosmological pipeline, and is not dependent on the details of the data-set investigated by the cosmological pipeline. As such, we test our methodology on a simple, simulated dataset which mimics the 3 year DES data-set~\citep{BroutSako2019}, including the cadence\added{, spectroscopic selection, }and observational noise~\citep{DAndreaSmith2018} of this data-set. We assume a flat, cold dark matter ($wCDM$) cosmology with $H_{0}=70$km/s/Mpc, $\Omega_{M}=0.3$, and $w=-1.0$. The DES 3 year data-set includes a previously released low-$z$ sample from several sources. We simplify the low-$z$ simulation by generating a DES-like sample for $0.0\le{}z\le{}0.08$ with the same statistics as the low-$z$ sample. Additionally, we only simulate SNe Ia and do not consider contamination from core-collapse supernovae, so that we can keep our analysis as simple as possible. The true DES data-set includes data from a variety of telescopes, as well as misclassified core-collapse SNe, so if our methodology were to be used to test the DES analysis, these details will need to be included in all simulations. \added{The redshift distribution of our DES and low-$z$ simulated sample is presented in Figure~\ref{fig:redshift}. An example of a simulated lightcurve is presented in Figure~\ref{fig:lightcurve}.}

\begin{figure}
    \centering
    \includegraphics[width=\linewidth]{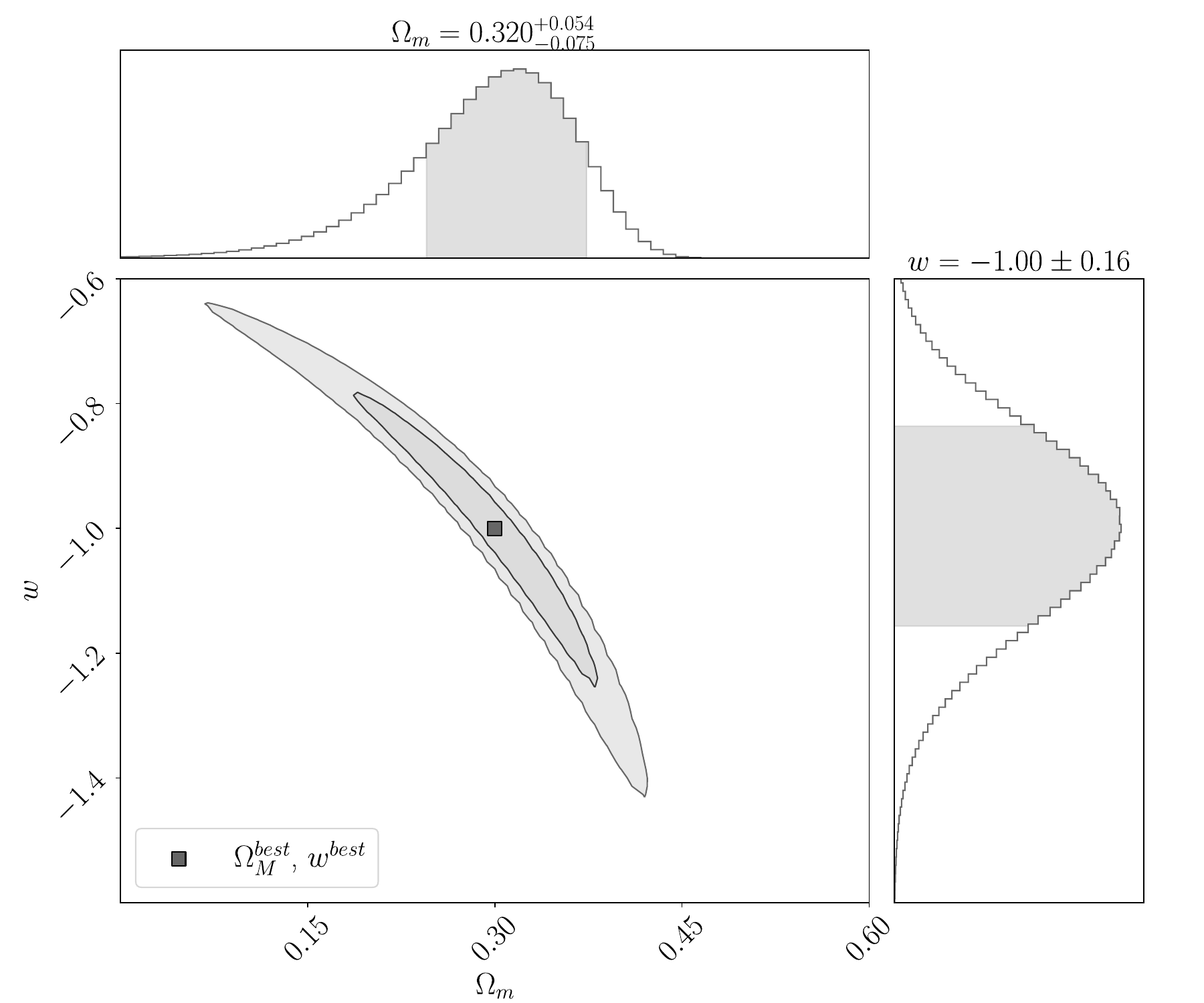}
    \caption{The cosmological contour produced by Pippin for our simulated data-set. The aim of our methodology is to test the consistency of this contour. The central panel shows the 2-D 68\% and 95\% contours, whilst the top and right panel show the marginalised, 1-D contour for $\Omega_{M}$ and $w$ respectively. Here $\Omega_{M}^{best}=0.320^{+0.054}_{-0.075}$ and  $w^{best}=-1.00\pm0.16$}
    \label{fig:Contour}
\end{figure}

We analyse this simulated data-set with Pippin in order to produce the cosmological contour \replaced{that we aim to validate}{we check} (shown in Figure~\ref{fig:Contour}), and to calculate the best fitting cosmology:
\begin{align}\label{eq:best}
    \begin{split}
        \Omega_{M}^{best} &= 0.32^{+0.054}_{-0.075} \\
        w^{best} &= -1.00\pm0.16
    \end{split}
\end{align}

\section{Methods}\label{sec:methods}
Here we describe our methodology for estimating the confidence region of a cosmological data-set using Neyman construction. To aid the reader, a glossary of terms used throughout is provided in Table~\ref{tab:glossary}.

\begin{table}
    \centering
    \begin{tabular}{p{0.15\linewidth}|p{0.75\linewidth}}
        Parameter & Description  \\
        $\Omega_{M}^{best}$, $w^{best}$ & The best-fitting output cosmology for our experiment data-set. Described in Section~\ref{sec:experiment}. \\
        $\Omega_{M}'$, $w'$ & A strategically selected input cosmology from which 150 realisations will be drawn. Described in Section~\ref{sec:percentile}. \\
        $\vec{\Omega_{M}}$, $\vec{w}$ & A distribution of 150 best-fitting cosmologies, produced by processing the 150 realisations of $\Omega_{M}'$, $w'$ with Pippin. Described in Section~\ref{sec:percentile}. \\
        $w^{*}(\Omega_{M})$ & A one dimensional function that approximates $\vec{\Omega_{M}}$, $\vec{w}$. Found by fitting a Gaussian process through $\vec{\Omega_{M}}$, $\vec{w}$. Described in Section~\ref{sec:ellipse}.
    \end{tabular}
    \caption{A glossary of terms used in our methodology, which are defined throughout the text}
    \label{tab:glossary}
\end{table}

For a given input cosmology, the Neyman construction provides a prescription for using simulations to calculate the percentile contour, or the boundary of a confidence region, that input cosmology lies on. By calculating these percentile contours for a grid of input cosmologies, the confidence region can be estimated as the set of input cosmologies which lie on percentile contours less than or equal to the desired confidence level. The extensive compute time of supernova simulations makes it difficult to densely sample the parameter space, so we instead create an approximate Neyman construction by strategically choosing a small number of input cosmologies at representative locations on the 68\% contour.

In Section~\ref{sec:percentile} we describe how to calculate the percentile contour for a single cosmological input. In Section~\ref{sec:confidence_region} we describe how we find cosmological inputs which lie on the 68\% percentile contour, and how we estimate the confidence region.

\subsection{Calculating the percentile contour}\label{sec:percentile}
\begin{figure}
    \centering
    \includegraphics[width=\linewidth]{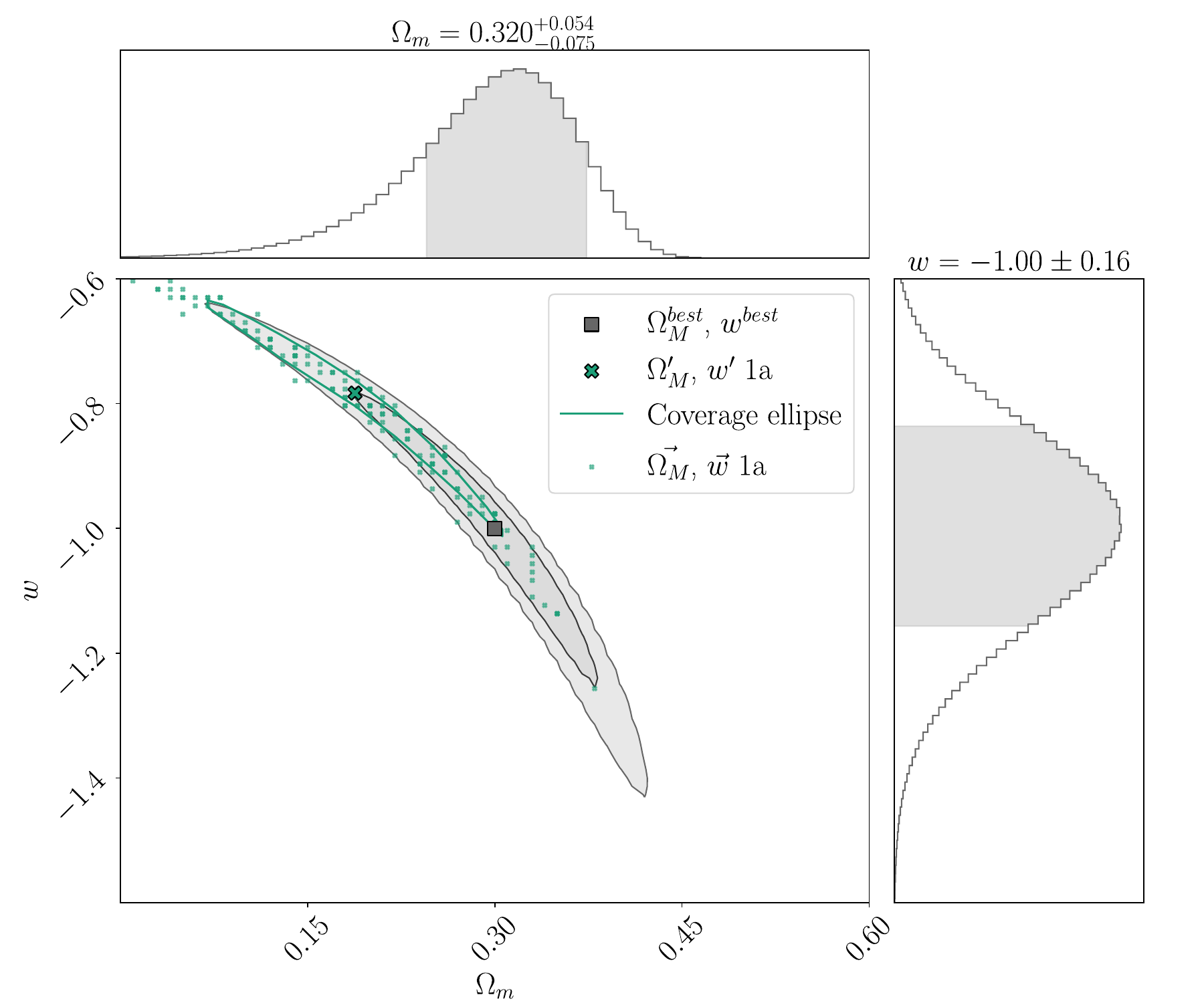}
    \caption{An example of using simulations to calculate the percentile contour for $\Omega_{M}'$, $w'$, where $\Omega_{M}^{best}$, $w^{best}$ represent the best fitting cosmology for our test data-set. We simulate 150 data-sets using $\Omega_{M}'$, $w'$ as the input, and process each data-set with Pippin to find the best fitting cosmology. The coverage ellipse is defined to intersect $\Omega_{M}^{best}$, $w^{best}$. The percentile contour for $\Omega_{M}'$, $w'$ is the percentage of best fitting cosmologies contained within this coverage ellipse.}
    \label{fig:Likelihood}
\end{figure}

\begin{figure}
    \centering
    \includegraphics[width=\linewidth]{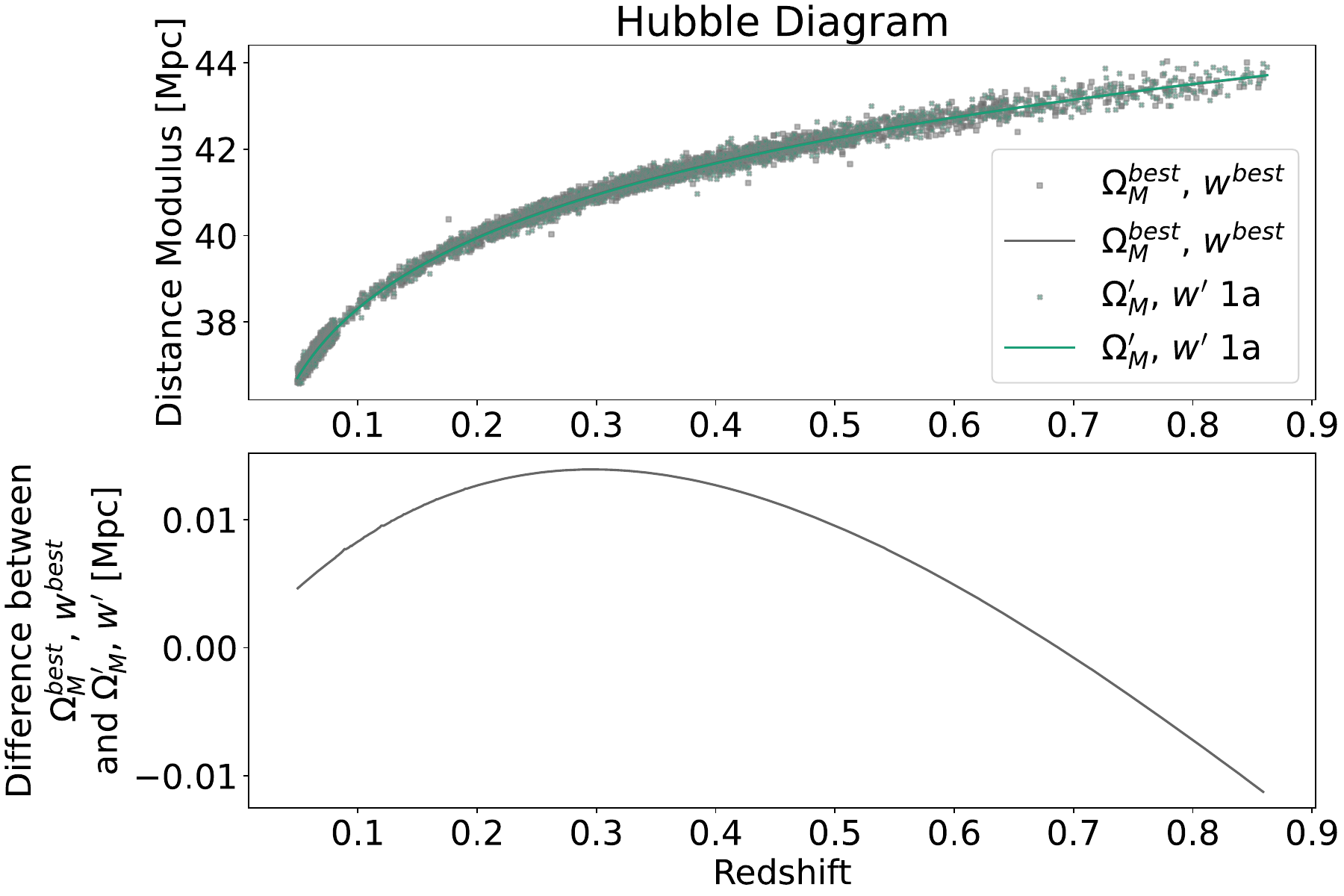}
    \caption{\added{Top Panel: Hubble Diagram for $\Omega_{M}^{best}=0.3$, $w^{best}=-1.0$ and $\Omega_{M}'=0.188$, $w'=-0.783$. This includes both simulated distance moduli and the analytic distance moduli based on the input cosmology.
    Bottom Panel: Difference between the analytic distance moduli of $\Omega_{M}^{best}$, $w^{best}$ and $\Omega_{M}'$, $w'$.}}
    \label{fig:Hubble}
\end{figure}

To calculate the percentile contour at any given cosmology ($\Omega_{M}'$, $w'$, which as noted above is chosen to lie at representative locations of the 68\% contour), we simulate 150 SNe Ia data-sets with $\Omega_{M}'$, $w'$ as the true cosmological input, using the procedure described in Section~\ref{sec:experiment}. \added{Each data-set is produced with a different random seed, allowing for statistical fluctuations between realisations.} We analyse each of these data-sets with Pippin to produce a distribution of best fitting cosmologies ($\vec{\Omega_{M}}$, $\vec{w}$) in the space of measured $\Omega_{M}$ and $w$. The Neyman construction predicts that the percentile contour for $\Omega_{M}'$, $w'$ is the percentage of these best fitting cosmologies encompassed by a coverage ellipse of this distribution. The coverage ellipse is defined to be centered on $\Omega_{M}'$, $w'$ and to intersect $\Omega_{M}^{best}$, $w^{best}$, which were calculated in Section~\ref{sec:experiment}, Equation~\ref{eq:best}. This ellipse represents the probability of a data-set with true cosmology $\Omega_{M}'$, $w'$ having a best fitting cosmology equal to $\Omega_{M}^{best}$, $w^{best}$. Figure~\ref{fig:Likelihood} shows an example of this calculation for $\Omega_{M}'=0.188$, $w'=-0.783$, which lies at one extreme end of the 68\% contour we are testing. \added{Figure~\ref{fig:Hubble} presents an example Hubble Diagram for both $\Omega_{M}^{best}$, $w^{best}$ and $\Omega_{M}'$, $w'$.}

\subsubsection{Fitting the coverage ellipse}\label{sec:ellipse}
The distribution of best fitting cosmologies about $\Omega_{M}'$, $w'$ usually follows the "banana" distribution that is typical of supernova cosmology, and is due to the inherent degeneracy between $\Omega_{M}$ and $w$. This contour shape makes it difficult to determine an accurate coverage ellipse around this distribution. To determine a coverage ellipse, we first fit a Gaussian Process (GP) through $\vec{\Omega_{M}}$, $\vec{w}$ to produce the one-dimensional function $w^{*}(\Omega_{M})$, which approximates $w$ as a function of $\Omega_{M}$ in the plane of $\vec{\Omega_{M}}$, $\vec{w}$. Next we subtract $w^{*}(\vec{\Omega_{M}})$ from $\vec{w}$ to transform the distribution of best fitting cosmologies into a more elliptical distribution. We fit a coverage ellipse to this transformed distribution which is centered on $\Omega_{M}'$, $w'$ and intersects $\Omega_{M}^{best}$, $w^{best}$.

\begin{figure}
    \centering
    \includegraphics[width=\linewidth]{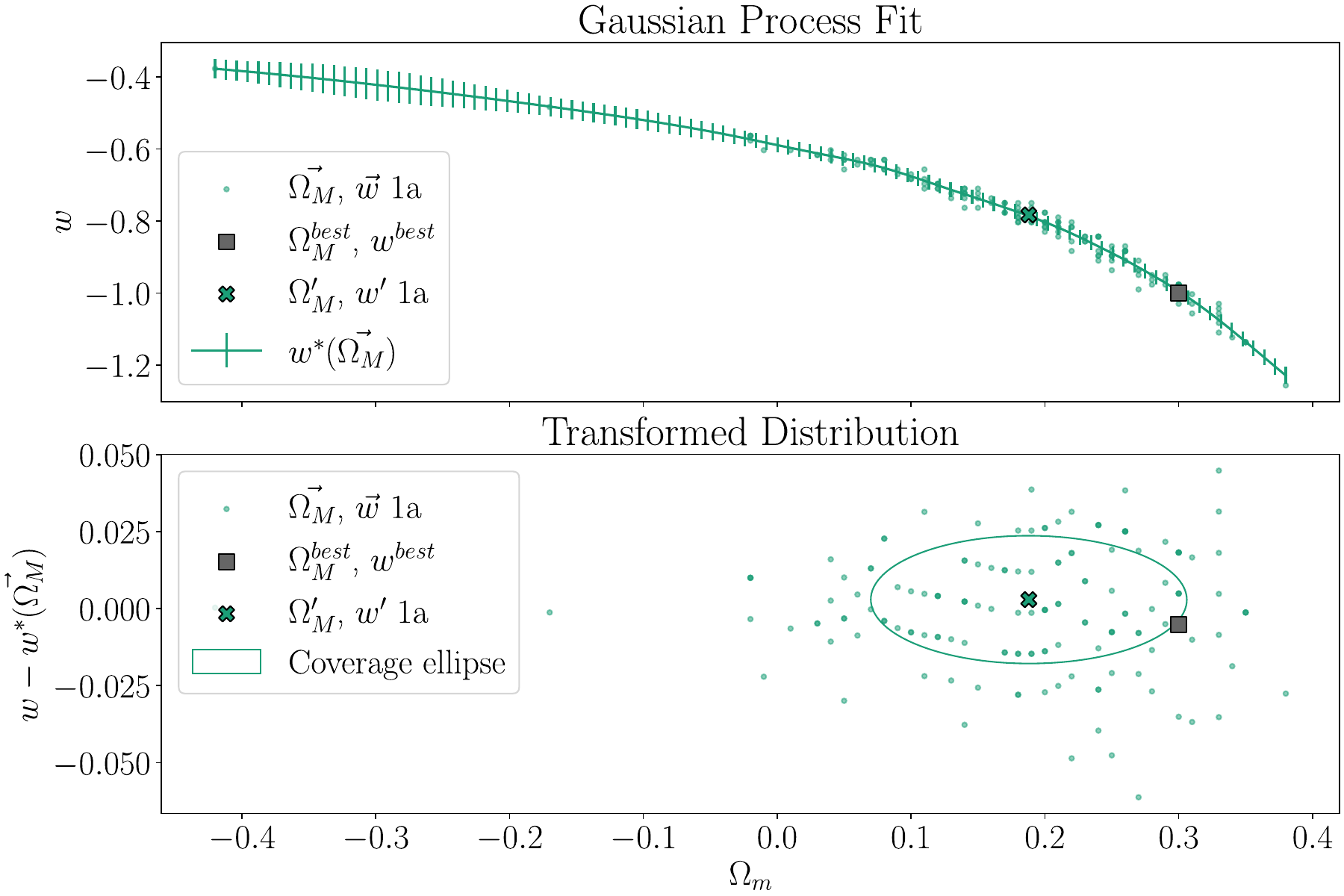}
    \caption{\textbf{Top panel:} A GP fit ($w^{*}(\vec{\Omega_{M}})$) to the best fitting output cosmologies of the 150 realisations with input cosmology: $\Omega_{M}'=0.188$, $w' = -0.783$. \textbf{Bottom panel:} The same distribution of maximum likelihood output cosmologies, transformed by subtracting $w^{*}(\vec{\Omega_{M}})$ from $\vec{w}$. This transformed distribution is more elliptical than the original distribution, and is more appropriate for fitting coverage ellipses. We show one such coverage ellipse in the bottom panel, scaled to intersect with the experiment cosmology input. In this example 46\% of the simulations are covered by the ellipse, so this $\Omega_{M}'$, $w'$ lies on the 46\% percentile contour.}
    \label{fig:GPE}
\end{figure}

The percentile contour for $\Omega_{M}'$, $w'$ is then the percentage of the best fitting cosmologies covered by this coverage ellipse. Figure~\ref{fig:GPE} shows an example of this transformation and ellipse fitting technique for $\Omega_{M}'=0.188$, $w'=-0.783$, the same cosmology as shown in Figure~\ref{fig:Likelihood}.

The uncertainty in the computed percentage is estimated by performing 1000 bootstrap resamples of the distribution of best fitting cosmologies, and typically results in an uncertainty in the percentile contour of $\pm4\%$.

\subsection{Estimating the confidence region}\label{sec:confidence_region}
We now have a statistically rigorous method of calculating the percentile contour over the cosmological parameter space. Using this method, we could compute a Neyman construction by computing the percentile contour across a grid that covers the cosmological parameter space, and from this determine a confidence region. However, each evaluation of the percentile contour requires 150 simulated data-sets, which is computationally expensive.
To more efficiently determine a confidence region, we develop an approximate Neyman construction method which requires far fewer simulations. Instead of evaluating the percentile contour over the entire cosmological parameter space, we use the bisection method to iteratively find the input cosmologies that lie on the 68\% percentile contour. These cosmologies define the edge of the 68\% confidence region.

\begin{figure}
    \centering
    \includegraphics[width=\linewidth]{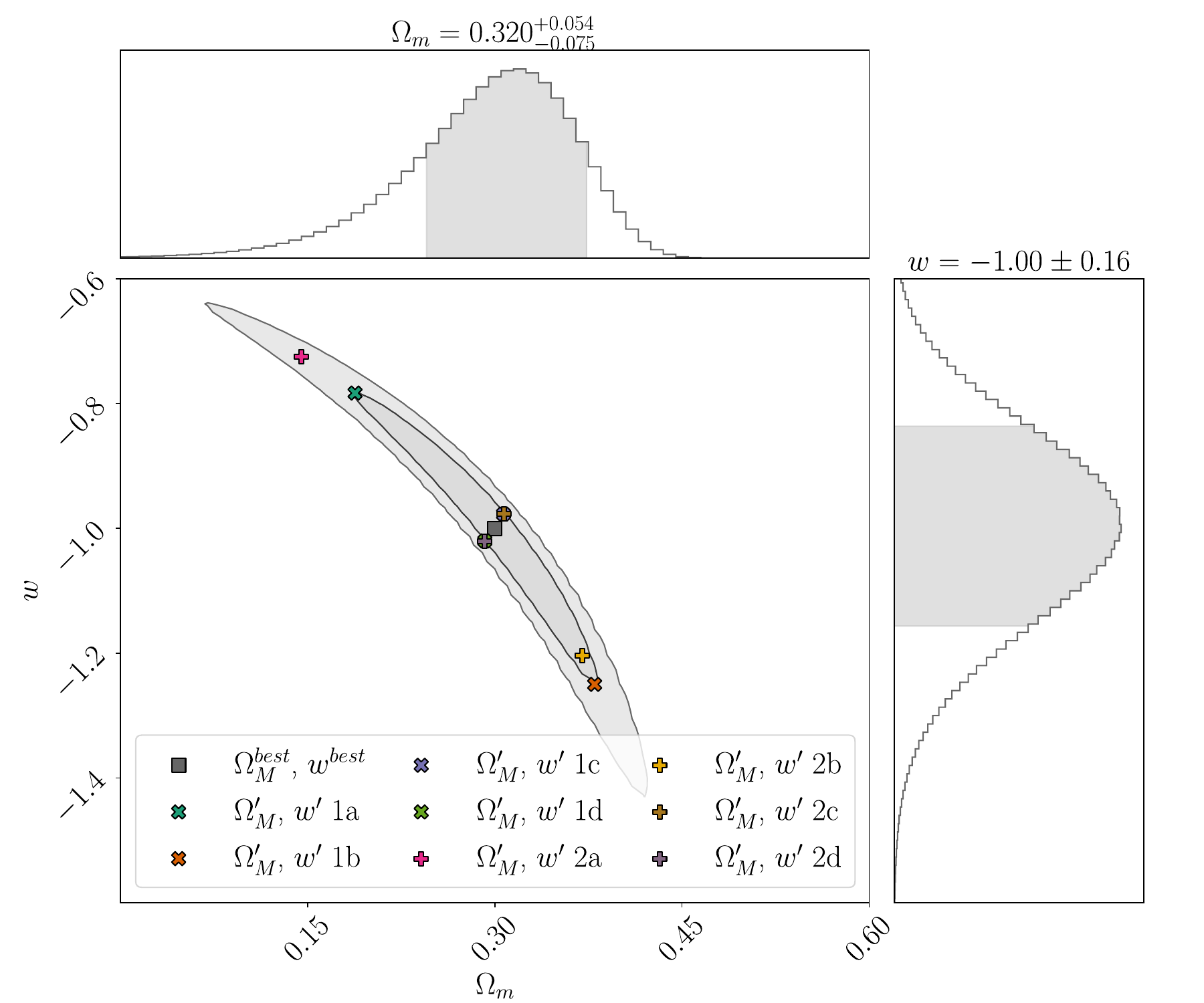}
    \caption{The input $\Omega_{M}$ and $w$ values for the experiment ($\Omega_{M}^{best}$, $w^{best}$) and the approximate Neyman construction ($\Omega_{M}'$, $w'$).}
    \label{fig:Neyman}
\end{figure}

We first calculate the percentile contour for cosmologies at several representative locations on the 68\% contour. We select two input cosmologies which are at the furthest extent of the 68\% contour, and two input cosmologies which are at the closest region of the 68\% contour to $\Omega_{M}^{best}$, $w^{best}$. This enables us to probe the consistency of the 68\% contour across its entire span. The cosmological inputs used in defining the approximate Neyman construction, and the percentile contour for each input are shown in Figure~\ref{fig:Neyman}, and detailed in Table~\ref{tab:InputPercentile}. Inputs 1a, 1b, 1c, and 1d were chosen to lie on the 68\% contour of the original experiment cosmology posterior. A second set of inputs (2a, 2b, 2c, and 2d) were chosen to compensate for how far the previous set of coverage ellipses differed from 68\% coverage, as described below.

If the input cosmologies lie at a percentile contour of $<68\%$ confidence, we select a new cosmology further from $\Omega_{M}^{best}$, $w^{best}$, and conversely select a cosmology closer to $\Omega_{M}^{best}$, $w^{best}$ if the initial percentile contour is $>68\%$ confidence. This allows us to find an input cosmology which lies on a percentile contour within one standard deviation of 68\%, as measured by bootstrap resampling. Though the iterative method typically converges to a percentile contour within one standard deviation of $68\%$ within 2 or 3 iterations, converging to exactly 68\% would take significantly more iterations. As such, once we are within one standard deviation of 68\%, we linearly interpolate or extrapolate to find an input cosmology which lies exactly on the 68\% percentile contour.

\begin{figure}
    \centering
    \includegraphics[width=\linewidth]{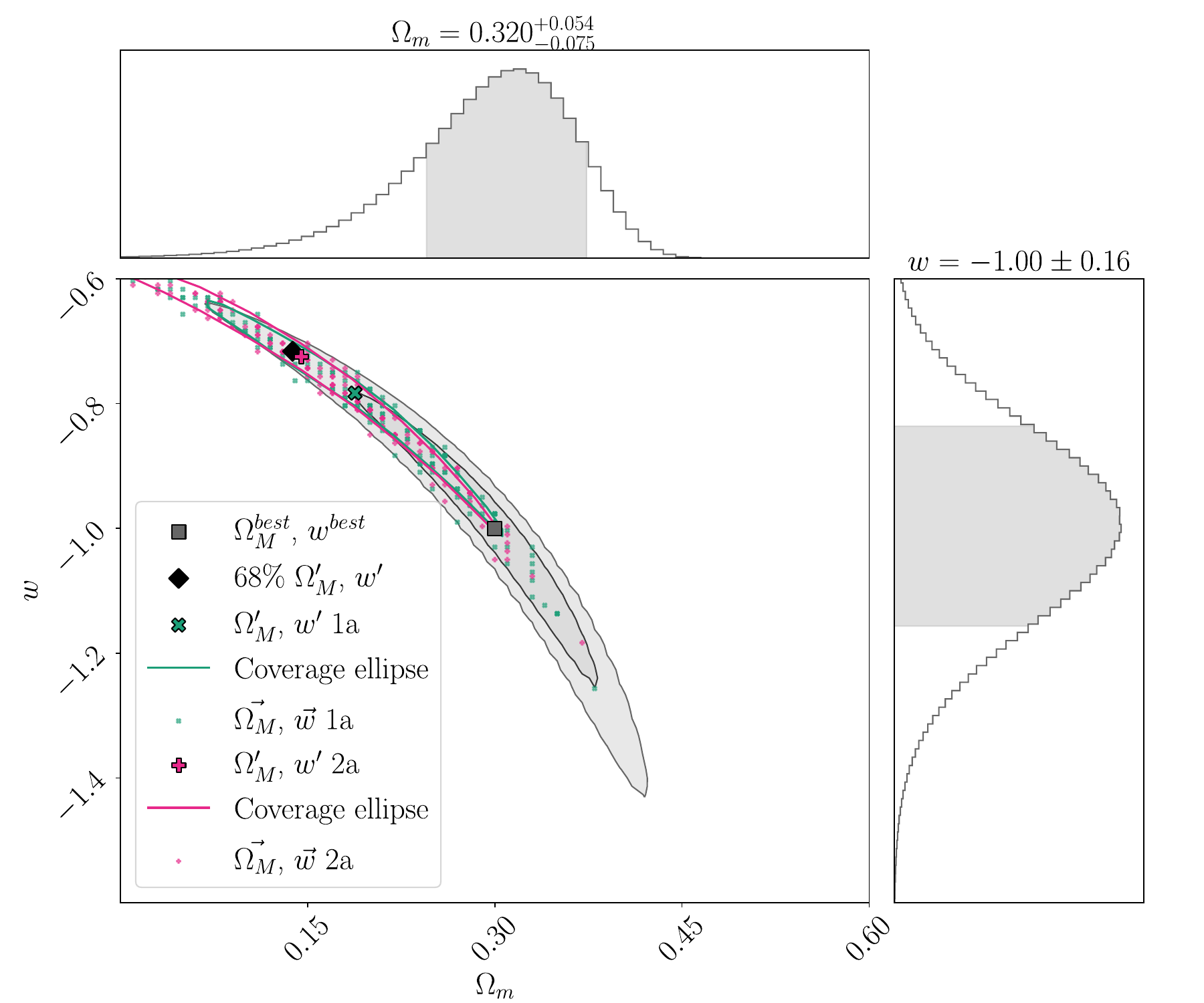}
    \caption{Example of finding the edge of the 68\% confidence region. $\Omega_{M}'$, $w'$ 1a was defined with an input cosmology on the extreme end of the Pippin 68\% contour, and was found to lie on the $46\%\pm4\%$ percentile contour. $\Omega_{M}'$, $w'$ $2a$ was found iteratively, and lies on the $65\%\pm4\%$ percentile contour. Linearly extrapolating from these two cosmological inputs gives us $68\%$ $\Omega_{M}'$, $w'$.}
    \label{fig:Approximate}
\end{figure}

\begin{figure*}
    \centering
    \includegraphics[width=\linewidth]{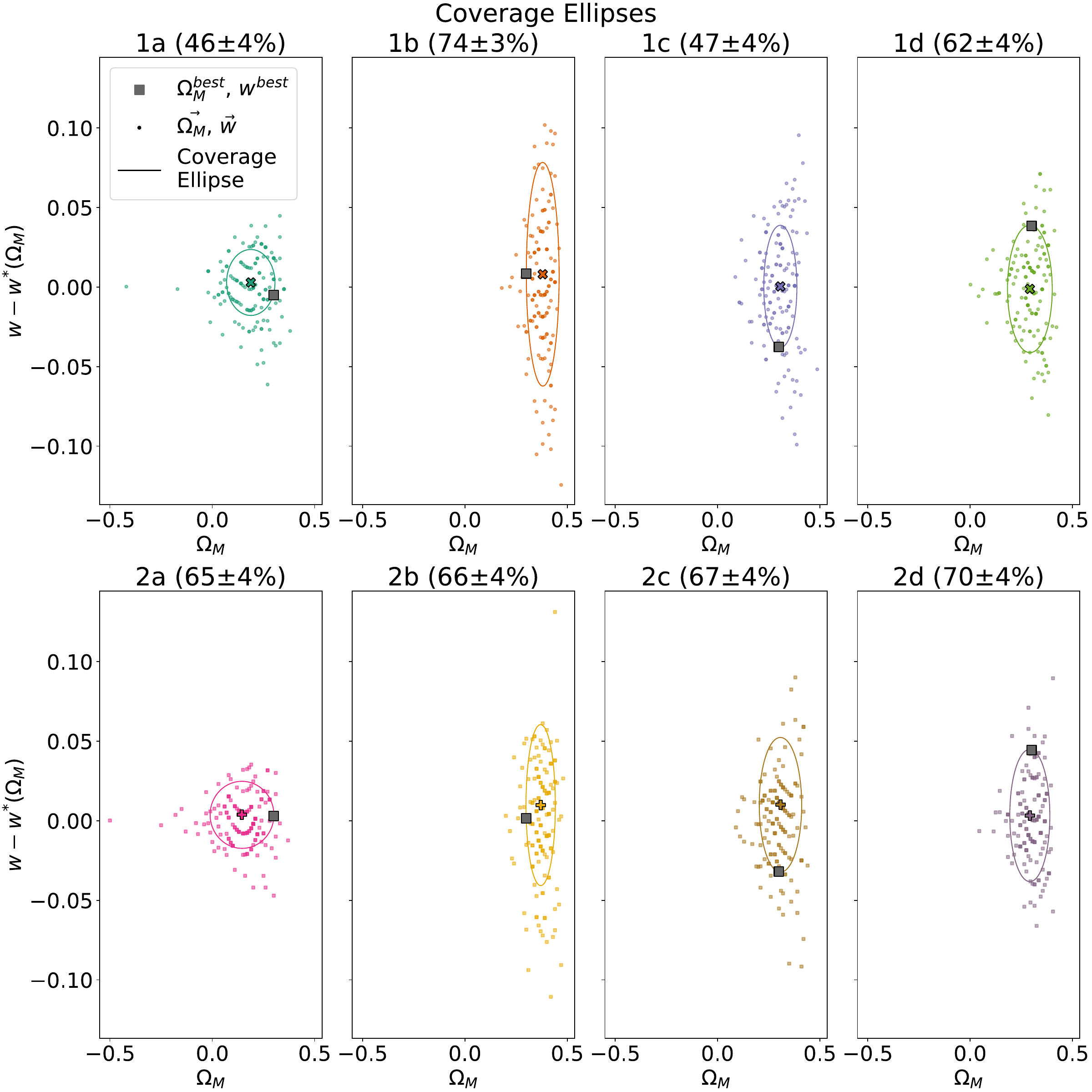}
    \caption{Coverage ellipses fit to the maximum likelihood distribution of each Neyman input cosmology, transformed to $\left\{\Omega_{m}, w-w^{*}(\Omega_{M})\right\}$. The coverage ellipse is defined to be centered on the Neyman input, scaled such that it contains the experiment cosmology input. The title of each plot shows the percentage of maximum likelihood output cosmologies covered by the ellipse, this is our numerical estimate of likelihood. The uncertainty in this estimate is calculated via 1000 bootstrap resamples and is 4\%.}
    \label{fig:Ellipse}
\end{figure*}

Figure~\ref{fig:Approximate} shows this approximate Neyman Construction technique for $\Omega_{M}'=0.188$, $w'=-0.783$, the same cosmology shown in Figures~\ref{fig:Likelihood} and~\ref{fig:GPE}. This cosmology was found to lie on the $46\%\pm4\%$ percentile contour, and after iteration the cosmology $\Omega_{M}'=0.145$, $w'=-0.725$ was found to lie on the $65\%\pm4\%$ percentile contour, which is within one standard deviation of 68\%. We linearly extrapolated from these two input cosmologies to find that $\Omega_{M}'=0.138$, $w'=-0.716$ lay on the 68\% percentile contour. Figure~\ref{fig:Ellipse} shows the coverage ellipses fit to each cosmological input.

\begin{table}
    \centering
    \begin{tabular}{l|l|l|l}
         Cosmology          & Input         & Input         & Percentile            \\
         Input              & $\Omega_{M}$  & $w$           & Contour               \\\hline
         Experiment         & 0.3           & -1.0          & -                     \\
         1a                 & 0.188         & -0.783        & $46\%\pm4\%$          \\
         1b                 & 0.38          & -1.25         & $74\%\pm4\%$          \\
         1c                 & 0.307         & -0.977        & $47\%\pm4\%$          \\
         1d                 & 0.292         & -1.02         & $62\%\pm4\%$          \\
         2a                 & 0.145         & -0.725        & $65\%\pm4\%$          \\
         2b                 & 0.37          & -1.204        & $66\%\pm4\%$          \\
         2c                 & 0.3075        & -0.9765       & $67\%\pm4\%$          \\
         2d                 & 0.2917        & -1.0205       & $70\%\pm4\%$
    \end{tabular}
    \caption{The input $\Omega_{M}$ and $w$ values for the experiment and approximate Neyman construction input cosmologies, as well as the percentile contour each cosmological input lies on.}
    \label{tab:InputPercentile}
\end{table}

\section{Results}\label{sec:results}
\begin{figure}
    \centering
    \includegraphics[width=\linewidth]{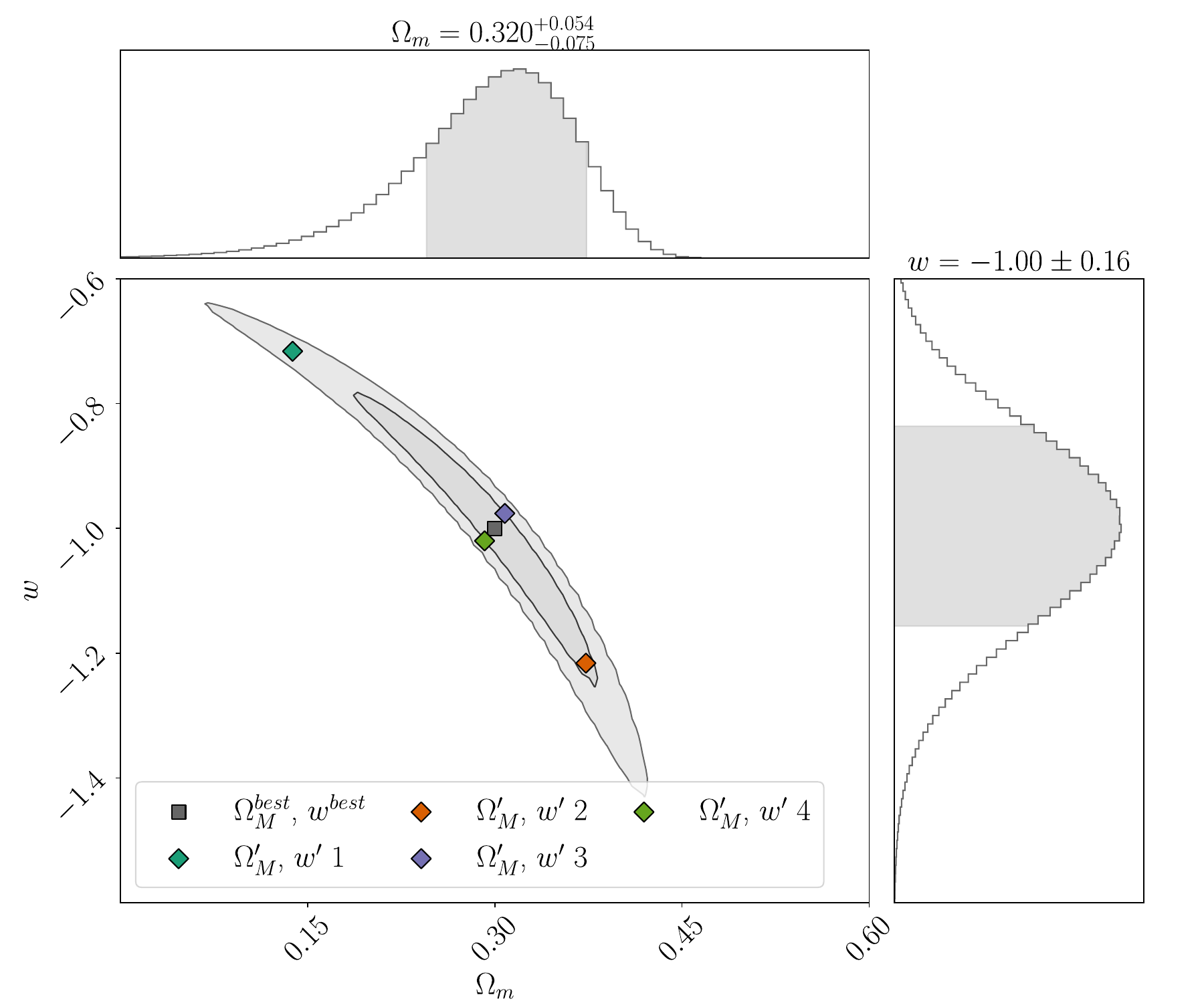}
    \caption{Comparison between the 68\% confidence region determined from our approximate Neyman construction, and the 68\% contour of the experiment cosmology. The confidence region is consistent with the contour close to the input cosmology, but displays an offset at the extreme ends of the contour. This offset is likely due to the bias correction method used by BBC, which is most accurate close to the input cosmology.}
    \label{fig:Final}
\end{figure}

In the previous section we describe how we computed the location of the 68\% confidence region at four strategically selected cosmological inputs in the $\Omega_{M}-w$ plane using an approximate Neyman construction. The results are presented in Table~\ref{tab:PercentDiff} and shown in Figure~\ref{fig:Final}.

We see the largest difference between the experiment cosmology 68\% contour and the 68\% confidence region in the first input cosmology, which lies in the top left quadrant of the cosmological contour. We find a shift of $\sim$0.05 in $\Omega_{M}$ and a shift of $\sim$0.07 in $w$. The second input cosmology, which lies in the bottom right quadrant has a shift of $\sim$0.007 in $\Omega_{M}$ and $\sim$0.03 in $w$. These input cosmologies lie at the furthest extent of the 68\% cosmological contour from $\Omega_{M}^{best}$ and $w^{best}$. In contrast, the third and fourth input cosmologies, which lie much closer to the experiment input cosmology, have a shift of $\lesssim$0.001 in both $\Omega_{M}$ and $w$. The increase in discrepancy as we probe parameter space that is further from the experiment input cosmology is expected, \replaced{as the BBC bias correction is only performed at a single point in cosmological parameter space, the best-fitting cosmology. The bias correction depends on the input cosmology, thus the BBC bias correction produces a contour which is accurate close to the best-fitting cosmology, but induces an offset in the contour at parameter space further from the best-fitting cosmology.}{as the bias correction used by BBC is most accurate near the input cosmology, and performs a lower quality correction further from the input.}\added{By contrast, our Neyman construction method applies this bias correction in a cosmologically dependent manner across the entire parameter space, removing this offset.}

This offset is important to consider, especially when combining supernova contours with other cosmological probes such as the cosmic microwave background. Fortunately, it is the region of posterior space close to the input cosmology which overlaps with the contours of other cross-cutting probes. As such this offset is unlikely to significantly affect multi-probe cosmological analyses.

Where this offset could be significant is in the investigation of cosmological tensions, where accurate uncertainties \added{of the tails} are vital to successfully assess the significance of the tension. In these cases, it may be useful to use a method like our approximate Neyman construction to produce more accurate and statistically rigorous measure of the uncertainty in a cosmological fit.

\begin{table}
    \centering
    \begin{tabular}{l|l|l|l|l}
        Cosmology       & $\Omega_{M}'$     & $w'$      & $\Omega_{M}$ Absolute & $w$ Absolute  \\
        Input           &                   &           & Difference            & Difference    \\\hline
        1               & 0.138             & -0.716    & 0.05                  & 0.07          \\
        2               & 0.373             & -1.216    & 0.007                 & 0.03          \\
        3               & 0.308             & -0.976    & 0.001                 & 0.001         \\
        4               & 0.2918            & -1.02     & 0.0002                & 0.0
    \end{tabular}
    \caption{Comparison between the 68\% confidence region determined from our approximate Neyman construction, and the 68\% contour of the experiment cosmology. The absolute difference is the difference between the cosmologies at the edge of the 68\% contour produced by Pippin, and the cosmologies at the edge of the 68\% confidence region produced by our approximate Neyman construction.}
    \label{tab:PercentDiff}
\end{table}
\subsection{Effect of Bias Correction Input Cosmology}
\begin{figure}
    \centering
    \includegraphics[width=\linewidth]{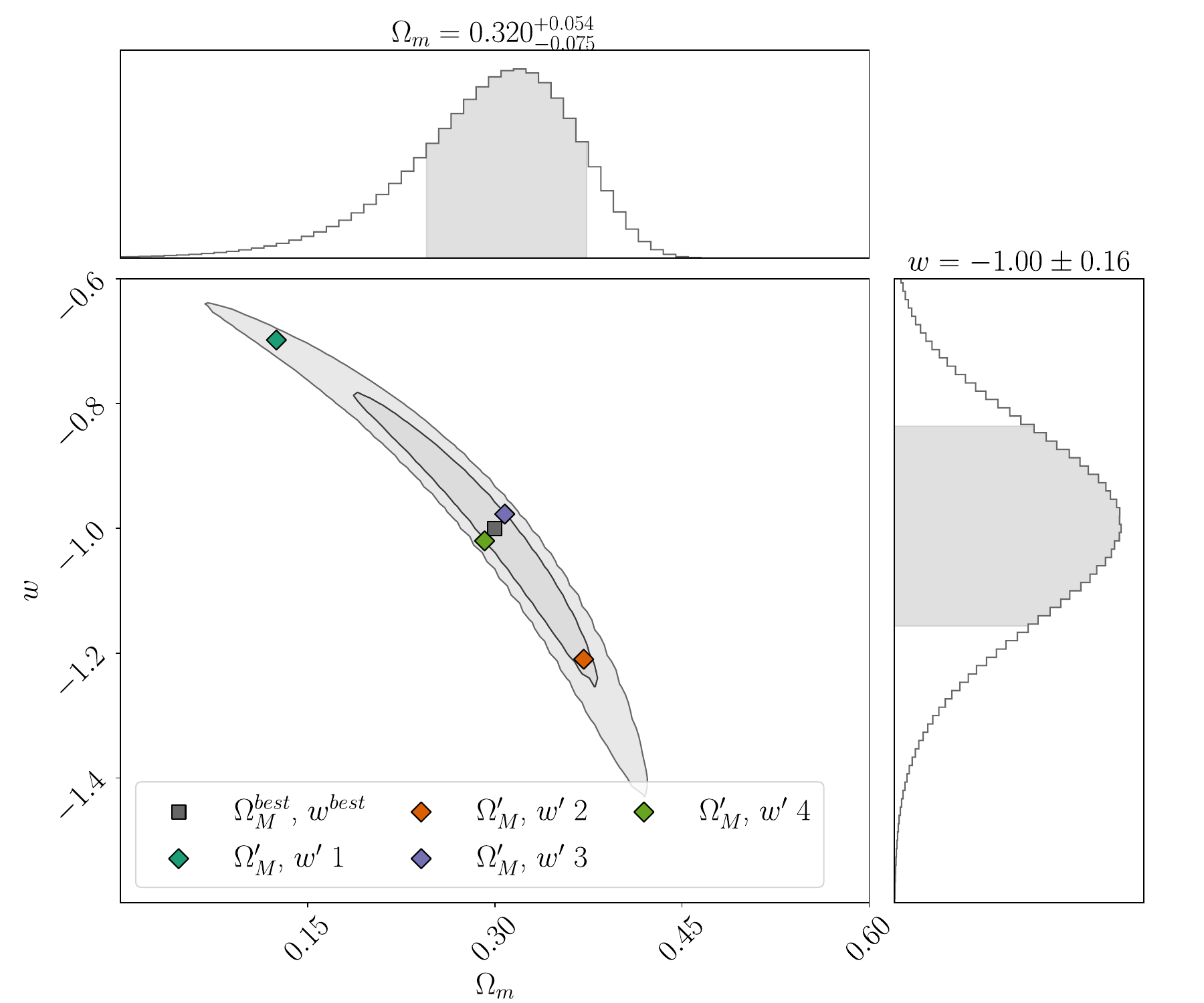}
    \caption{As per Figure~\ref{fig:Final}, but varying the bias correction simulation to match the input cosmology. Very similar results are found, indicating that the cosmology used for the bias correction is not significantly impacting the results.}
    \label{fig:BCFinal}
\end{figure}

We repeat our analysis with bias correction simulations that use a cosmology that is nearer to the data-set they are attempting to correct, rather than a single bias correction shared amongst all realisations. Figure~\ref{fig:BCFinal}, and Table~\ref{tab:BCPercentDiff} show the results of this repeat analysis. Our results are very similar to the case when we shared the bias correction simulation amongst all realisations, with the first and second cosmological input deviating the most. Overall, these results reinforce our suggestion that the offset present between the cosmological contour and confidence region is caused by the BBC bias correction, however the choice of bias correction does not significantly affect our consistency test. If computational cost is a concern, using only one bias correction simulation shared amongst all realisations will significantly reduce the computational cost of our approximate Neyman construction method, without significantly reducing the quality of the consistency test.
\begin{table}
    \centering
    \begin{tabular}{l|l|l|l|l}
         Cosmology  & $\Omega_{M}'$      & $w'$      & $\Omega_{M}$ Absolute     & $w$ Absolute      \\
         Input      &                    &           & Difference                & Difference        \\\hline
         1          & 0.125              & -0.698    & 0.063                     & 0.085             \\
         2          & 0.371              & -1.21     & 0.009                     & 0.004             \\
         3          & 0.308              & -0.977    & 0.001                     & 0.0               \\ 
         4          & 0.292              & -1.02     & 0.0                       & 0.0
    \end{tabular}
    \caption{As for Table~\ref{tab:PercentDiff}, but varying the bias correction simulation to match the input cosmology.}
    \label{tab:BCPercentDiff}
\end{table}%

\section{Conclusions}\label{sec:conclusions}
In this paper we present a statistically rigorous method for checking the consistency of contours produced in a  cosmological analysis. To achieve this, we implement an approximate Neyman construction which requires far less computation than a true Neyman construction. This approximate Neyman construction is then used to define the 68\% confidence region for a single cosmological realisation. We use this confidence region to test the consistency of the 68\% contour produced by the BBC framework, as integrated in Pippin, although this method can be used to test the consistency of any cosmological parameter estimation method. This represents the first time the BBC framework has been tested with a statistically rigorous methodology. 

Our analysis showed that, for a DES-3YR like dataset, Pippin is producing reasonable, consistent parameter estimates. There was some discrepancy between the CR and the cosmological contour when considering the farthest extent of the 68\% contour. This discrepancy was, at maximum, a shift of $\sim$0.05 in $\Omega_{M}$, and $\sim$0.07 in $w$, and was likely due to the accuracy of BBC's bias correction being best when close to the input cosmology, and degrading in regions of parameter space which are far from the input cosmology. It is also important to recognise that this does not correspond to an equivalent error in the reported maximum posterior cosmological parameters. When considering cosmological inputs close to the experiment cosmology input, the confidence region and cosmological contour had near perfect agreement. As such any overall discrepancy is unlikely to significantly effect the results of a cosmological analysis, especially when multiple cross-cutting probes are combined. However, this shift is important when considering analyses concerned with assessing cosmological tensions - where the precise shape and size of the contour are vitally important to the analysis.

We see very similar results when each realisation had its own bias correction simulation, rather than sharing one bias correction simulation amongst all realisations, indicating that a sensible choice of bias correction is not likely to significantly effect our consistency checks.

Overall, we believe our method for consistency checking cosmological contours with an approximate Neyman construction represents an important improvement in the statistical rigour applied to cosmological analyses, and should \replaced{become a standard step in all cosmological analyses}{be considered carefully when a survey analysis is performed}. Our methodology can also be used to rigorously test cosmological contours for other cosmological probes, which have similarly complex pipelines. We believe this method will be particularly useful for future analyses, such as the DES 5-year supernova analysis, and the upcoming LSST survey. We plan to repeat this analysis using simulations that match the DES 5-year supernova analysis to test the consistency of those results.%

\section{Acknowledgments}\label{sec:acknowledgments}
This is a pre-copyedited, author-produced PDF of an article accepted for publication in the Publications of the Astronomical Society of Australia following peer review. The version of record~\citep{} is available online at: \url{}.

We acknowledge parts of this research were carried out on the traditional lands of the Ngunnawal, Turrbal, and Jagera people.  We pay our respects to their elders past, present, and emerging.
P.A. was supported by an Australian Government Research Training Program (RTP) Scholarship.
H.Q. and M.S. were supported by DOE grant DE-FOA-0002424 and NSF grant AST-2108094.
B.E.T. and his group  were supported by the Australian Research Council Centre of Excellence for All Sky Astrophysics in 3 Dimensions (ASTRO 3D), through project number CE170100013.  T.M.D. was supported by an Australian Research Council Laureate
Fellowship FL180100168.%
\subsection{Data availability}
The data underlying this article are available in the article and in its online supplementary material. The code used throughout this article is available at~\url{https://github.com/dessn/BiasValidation}.
\subsection{Software}\label{sec:software}
Pippin~\citep{HintonBrout2020}, SNANA~\citep{KesslerBernstein2009}, ChainConsumer~\citep{chainconsumer}, Numpy~\citep{numpy}, MatPlotLib~\citep{matplotlib}, SciKit-Learn~\citep{scikit-learn}, Scipy~\citep{scipy}%
\subsection{Facilities}\label{sec:facilities}
This work was completed in part with resources provided by the University of Chicago Research Computing Center.%
\pagebreak
\bibliography{biblio.bib}

\printendnotes

\end{document}